\documentclass[a4paper,12pt]{article}

\usepackage{amsmath}
\usepackage[psamsfonts]{amssymb}
\usepackage{rsfs}
\usepackage{bm}

\usepackage{cite}
\usepackage[dvips]{graphicx}
\usepackage{color}

\begin{document} 

\begin{titlepage}

\baselineskip 10pt
\hrule 
\vskip 5pt
\leftline{}
\leftline{Chiba Univ. Preprint
          \hfill   \small \hbox{\bf CHIBA-EP-153}}
\leftline{\hfill   \small \hbox{hep-th/0504198}}
\leftline{\hfill   \small \hbox{May 2005}}
\vskip 5pt
\baselineskip 14pt
\hrule 
\vskip 1.0cm
\centerline{\Large\bf 
BRST symmetry of 
}
\vskip 0.4cm
\centerline{\Large\bf  
SU(2) Yang-Mills theory
}
\vskip 0.4cm
\centerline{\Large\bf  
in Cho--Faddeev--Niemi decomposition
}

\vskip 0.5cm

\centerline{{\bf 
K.-I. Kondo$^{\dagger,\ddagger,{1}}$,  
T. Murakami$^{\ddagger,{2}}$,
and 
T. Shinohara$^{\ddagger,{3}}$
}}  
\vskip 0.5cm
\centerline{\it
${}^{\dagger}$Department of Physics, Faculty of Science, 
Chiba University, Chiba 263-8522, Japan
}
\vskip 0.3cm
\centerline{\it
${}^{\ddagger}$Graduate School of Science and Technology, 
Chiba University, Chiba 263-8522, Japan
}
\vskip 1cm

\begin{abstract}
We determine the nilpotent BRST and anti-BRST transformations for the Cho--Faddeev-Niemi variables for the SU(2) Yang-Mills theory based on the new interpretation given in the previous paper of the Cho--Faddeev-Niemi decomposition.  
This gives a firm ground for performing the BRST quantization of the Yang--Mills theory written in terms of the Cho--Faddeev-Niemi variables. 
We propose also a modified version of the new Maximal Abelian gauge which could play an important role in the reduction to the original Yang-Mills theory.

\end{abstract}


\hrule  
\vskip 0.1cm
${}^1$ 
  E-mail:  {\tt kondok@faculty.chiba-u.jp}
  
${}^2$ 
  E-mail:  {\tt tom@cuphd.nd.chiba-u.ac.jp}
  
${}^3$ 
  E-mail:  {\tt sinohara@cuphd.nd.chiba-u.ac.jp}

\par 
\par\noindent


\vskip 0.5cm

\newpage
\pagenumbering{roman}




\end{titlepage}


\pagenumbering{arabic}

\baselineskip 14pt
\section{Introduction}
\setcounter{equation}{0}

The change of variables is known to be a useful method of extracting the most relevant degrees of freedom to the physics in question. 
A proposal along this direction for the topological degrees in Yang-Mills theory was first made by Cho \cite{Cho80}, which have recently been readdressed by Faddeev and Niemi \cite{FN99}, and another by Faddeev and Niemi \cite{FN02}.  
The former is called the Cho--Faddeev--Niemi (CFN) decomposition in the literatures, while the latter is called the Faddeev--Niemi decomposition.
We focus on the former in this paper. 

In applying the CFN decomposition, however, there were much controversy \cite{Shabanov99,Gies01} over the treatment and the interpretation. 
A lot of progress toward the resolution of the conceptual issues was already made in \cite{BCK01,Cho03}, while the CFN decomposition was applied to reveal various non-perturbative features overlooked so far \cite{FN97,LN99,Gies01,BCK01,FN02,Cho03,Kondo04}. For example,   
 the Skyrme--Faddeev model \cite{FN97} describing glueball as the knot soliton solution may be deduced from the Yang-Mills theory by way of the CFN decomposition \cite{LN99,Gies01,BCK01,FN02,Cho03,Kondo04}. 
Moreover, the instability of the Savvidy vacuum \cite{Savvidy77} disappears by eliminating a tachyon mode \cite{NO78}, see \cite{Cho03} for the massless gluon and \cite{Kondo04,KKMSS05} for the massive gluon caused by a novel magnetic condensation.

In the previous paper \cite{KMS05}, we have given a new interpretation of the Yang-Mills theory written in terms of the CFN variables, called the CFN-Yang-Mills theory. This interpretation enables us to elucidate how the CFN-Yang-Mills theory with the enlarged local gauge symmetry is reduced to the gauge theory with the same local and global gauge symmetries (Yang-Mills theory II) as the original Yang-Mills theory (Yang-Mills theory I).  In fact, this is achieved by imposing the new version of the Maximal Abelian gauge (new MAG), which plays quite a different role from the conventional MAG \cite{tHooft81,KLSW87}. The new interpretation disposes of a pending question of discrepancy between the CFN-Yang-Mills theory and the original Yang-Mills theory for independent degrees of freedom. 

In the paper \cite{KKMSS05}, we have already discussed how to perform the numerical simulations based on this interpretation and we have actually performed the Monte Carlo simulations on a lattice. 
This is the first implementation of the CFN decomposition on a lattice. 

In this paper, moreover, we will discuss how to perform the BRST quantization of the CFN-Yang--Mills theory in the continuum formulation, as announced in \cite{KMS05}. 
For this purpose, we must first determine the nilpotent BRST transformations for  the ghost-antighost fields and the Nakanishi-Lautrup auxiliary fields,  in addition to those for the original CFN variables.  
Here it should be remarked that we must introduce more ghost and antighost fields reflecting the enlarged local gauge symmetry of the CFN-Yang--Mills theory. These BRST transformations are determined independently from the gauge fixing condition, i.e., the new MAG, as usual.  

In order to fix the whole gauge degrees of the CFN-Yang-Mills theory, we must impose one more gauge-fixing condition, e.g., Lorentz covariant gauge $\partial^\mu \mathscr{A}_\mu=0$, in addition to the new MAG.  
The additional gauge fixing is necessary to fix the local gauge symmetry II in the Yang-Mills theory II which is obtained by taking the new MAG from the CFN-Yang-Mills theory. 
Then, we can obtain the explicit form of the Faddeev--Popov (FP) ghost terms associated to two gauge-fixing conditions, once the additional gauge-fixing condition is specified. 
Therefore, we can uniquely determine the total Lagrangian, based on which the BRST quantization is performed. 

Moreover, we introduce another  gauge transformation and the corresponding BRST transformation, called the primed transformations.  This gives an alternative but equivalent description of the gauge and BRST symmetries of the CFN-Yang--Mills theory. 
Finally, we  determine also the nilpotent anti-BRST transformation for the CFN variables. 
As an application,  we propose a modified version of the new MAG, which is both BRST and anti-BRST exact simultaneously.



\section{Yang-Mills theory in the CFN decomposition}
\setcounter{equation}{0}


\subsection{Local gauge symmetry in terms of the CFN variables}


The Cho--Faddeev-Niemi (CFN) decomposition (or change of variables) of the original Yang-Mills gauge field $\mathscr{A}_\mu(x)$ is performed as follows. 
We restrict our consideration to the gauge group $G=SU(2)$.
First, we introduce a unit vector field $\bm{n}(x)$, i.e.,
\begin{align}
  \bm{n}(x) \cdot \bm{n}(x) := n^A(x) n^A(x) = 1 \quad (A=1,2,3) . 
  \label{nn=1}
\end{align}
Then the CFN decomposition is written in the form,
\begin{equation}
\mathscr{A}_\mu(x)
 =c_\mu(x) \bm{n}(x)
  +g^{-1}\partial_\mu \bm{n}(x)\times \bm{n}(x)
  +\mathbb X_\mu(x) ,
\end{equation}
where $\mathbb{X}_\mu(x)$ is perpendicular to $\bm{n}$,
\begin{align}
  \bm{n}(x) \cdot \mathbb{X}_\mu(x) = 0 .
  \label{nX=0}
\end{align}
For later convenience, we introduce
\begin{align}
    \mathbb{V}_\mu(x)  
    = c_\mu(x) \bm{n}(x)
  +g^{-1}\partial_\mu \bm{n}(x)\times \bm{n}(x) .
\end{align}

The first important observation made in \cite{KMS05} is that the restricted potential $c_\mu$ and gauge covariant potential $\mathbb{X}_\mu$ are specified by $\bm{n}$ and $\mathscr{A}_\mu$ as 
\begin{align}
c_\mu(x)
 &={\bm n}(x)\cdot\mathscr{A}_\mu(x), 
\label{def:c}
\\
\mathbb X_\mu(x)
 &=g^{-1}{\bm n}(x)\times D_\mu[\mathscr{A}]{\bm n}(x) ,
\label{def:X}
\end{align}
and, therefore, {\it the local gauge transformations $\delta c_\mu$ and $\delta\mathbb X_\mu$ are uniquely determined, once  only the transformations 
$\delta{\bm n}$ and $\delta\mathscr{A}_\mu$ are specified}. 
This fact indicates a special role played by the $\bm{n}$ field in the gauge transformation.

In the previous paper \cite{KMS05}, we have considered the local gauge symmetry respected by the Yang-Mills theory expressed in terms of the CFN variables, which we call {\it CFN--Yang-Mills theory} for short.  
We have shown \cite{KMS05} that 
 {\it the CFN--Yang-Mills theory has the local gauge symmetry 
 $SU(2)_{local}^{\omega} \times [SU(2)/U(1)]_{local}^{\theta}$ which is larger than the local SU(2) symmetry of the original Yang-Mills theory},
  since we can rotate the CFN variable $\bm{n}(x)$ by angle $\bm{\theta}^{\perp}(x)$ independently of the gauge transformation parameter $\bm{\omega}(x)$ of  $\mathscr{A}_\mu(x)$.  
Here the local gauge transformations of the CFN variables are given by 
\begin{subequations}
\begin{align}
\delta\mathscr{A}_\mu(x)
  =& D_\mu[\mathscr{A}]{\bm\omega}(x) ,
  \label{gtA}
\\
\delta{\bm n}(x)
  =& g{\bm n}(x) \times {\bm\theta}(x) 
  =g{\bm n}(x) \times {\bm\theta}_\perp(x) ,
  \label{gtn}
\\
  \delta c_\mu(x) 
  =&  g(\bm{n}(x) \times \mathscr{A}_\mu(x)) \cdot (\bm{\omega}_\perp(x) - \bm{\theta}_\perp(x)) + \bm{n}(x) \cdot \partial_\mu \bm{\omega}(x) ,
  \label{gtc}
\\
  \delta \mathbb{X}_\mu(x) 
  =& g \mathbb{X}_\mu(x) \times  (\bm{\omega}_\parallel(x)+\bm{\theta}_\perp(x)) + D_\mu[\mathbb{V}](\bm{\omega}_\perp(x)-\bm{\theta}_\perp(x))  ,
  \label{gtX}
\end{align}
\end{subequations}
where ${}_\parallel$ and ${}_\perp$ denote the parallel and perpendicular part to $\bm{n}$, 
and we have applied the gauge transformation (\ref{gtA}) and (\ref{gtn}) to (\ref{def:c}) and (\ref{def:X}) to obtain (\ref{gtc}) and (\ref{gtX}).  
If $\bm\omega_\perp(x)=\bm\theta_\perp(x)$,
the transformation (\ref{gtc}) and (\ref{gtX}) reduce to the gauge transformation II \cite{Kondo04,KMS05} with the parameter $\bm\omega'(x)=(\bm\omega_\parallel(x),\bm\omega_\perp(x)=\bm\theta_\perp(x))$.
Therefore,  the gauge transformation II corresponds to a special case 
$\bm\omega_\perp(x)=\bm\theta_\perp(x)$.
\underline{Local gauge transformation II}:  
\begin{subequations}
\begin{align}
  \delta_{\omega}' \bm{n}  =& g \bm{n} \times \bm{\omega'}  ,
\\
 \delta_{\omega}' c_\mu =&    \bm{n} \cdot \partial_\mu \bm{\omega'}   ,
\\
  \delta_{\omega}' \mathbb{X}_\mu =&  g \mathbb{X}_\mu \times \bm{\omega'} ,
\\
\Longrightarrow  & 
\delta_{\omega}' \mathbb{V}_\mu =   D_\mu[\mathbb{V}] \bm{\omega'}    . 
\end{align}
\end{subequations}
This should be compared with
\footnote{
The gauge transformation I was called the passive or quantum gauge transformation, while II was called the active or background gauge transformation.  However, this classification is sometimes confusing and could lead to misleading results, since two gauge transformations I and II are not independent. 
}

\underline{Local gauge transformation I}:  
\begin{subequations}
\begin{align}
  \delta_\omega \bm{n}  =& 0  ,
\\
 \delta_\omega c_\mu =& \bm{n} \cdot  D_\mu[\mathscr{A}] \bm{\omega} ,
\\
  \delta_\omega \mathbb{X}_\mu =&     D_\mu[\mathscr{A}] \bm{\omega} - \bm{n}( \bm{n} \cdot  D_\mu[\mathscr{A}] \bm{\omega}) ,
\\ 
  \Longrightarrow  & 
  \delta_\omega \mathbb{V}_\mu 
=   \bm{n}( \bm{n} \cdot  D_\mu[\mathscr{A}] \bm{\omega})  . 
\end{align}
\end{subequations}


\subsection{A new interpretation of the CFN-Yang--Mills theory}


   In order to obtain the gauge theory with the same local gauge symmetry as the original Yang-Mills theory, therefore,  we proceed to impose a gauge fixing condition by which $SU(2)_{local}^{\omega} \times [SU(2)/U(1)]_{local}^{\theta}$ is broken down to $SU(2)$ which is a subgroup of $SU(2)_{local}^{\omega} \times [SU(2)/U(1)]_{local}^{\theta}$. 
In the previous paper \cite{KMS05}, we have found that a way to do this is to impose the minimizing condition 
\begin{align}
 0 = \delta\int d^D x \frac12\mathbb X_\mu^2 ,
 \label{MAGcond}
\end{align}
which we called the new Maximal Abelian gauge (new MAG).
This is shown as follows. 
Since  the relationship (\ref{def:X}) leads to 
$
   g^2 \mathbb X_\mu^2
=    (D_\mu[\mathscr{A}]{\bm n})^2 ,
$
the local gauge transformation of $\mathbb X^2$ is calculated as
\begin{align}
\delta\frac12\mathbb X_\mu^2
  = g^{-1}
     (D_\mu[\mathscr{A}]{\bm n}) \cdot 
     \{D_\mu[\mathscr{A}](\bm\omega_\perp - \bm\theta_\perp) \times \bm n \} ,
\label{eq:dX^2}
\end{align}
where we have used (\ref{gtn}) and (\ref{gtA}).
Therefore, the local gauge transformation II  does not change $\mathbb X^2$.
Then the average over the spacetime of  (\ref{eq:dX^2}) reads  
\begin{align}
\delta\int d^D x\frac12\mathbb X_\mu^2
  =- \int d^D x
     (\bm\omega_\perp-\bm\theta_\perp)\cdot
     D_\mu[\mathbb V]\mathbb X_\mu ,
 \label{minX2}
\end{align}
where we have used (\ref{def:X})  and integration by parts. 
Hence the minimizing condition (\ref{MAGcond}) 
for arbitrary $\bm\omega_\perp$ and  $\bm\theta_\perp$ 
yields a condition in the differential form:
\begin{equation}
\mathbb F_{\rm MA}
 = \bm{\chi} :=D_\mu[\mathbb V]\mathbb X_\mu
 \label{dMAG}
 \equiv0 .
\end{equation}
Note that (\ref{dMAG}) denotes two conditions, since 
$\bm{n} \cdot \bm{\chi} =0$ which follows from an identity 
$ \bm{n} \cdot  D_\mu[\mathbb V]\mathbb X_\mu = 0$. 
Therefore, the minimization condition (\ref{MAGcond}) works as a gauge fixing condition
except for the gauge transformation II, i.e., $\bm\omega_\perp(x)=\bm\theta_\perp(x)$.
For  
$\bm\omega_\perp(x)=\bm\theta_\perp(x)$, the condition (\ref{dMAG}) transforms covariantly, 
$
\delta \bm{\chi}
  = g \bm{\chi} \times  (\bm{\omega}_\parallel +\bm{\omega}_\perp ) 
  = g \bm{\chi} \times  \bm{\omega}  .
$
Here the local rotation of $\bm{n}$,  
$
\delta{\bm n}(x)  =g{\bm n}(x) \times {\bm\theta}_\perp(x)
$,
leads to  $\delta \bm{\chi}=0$ on $\bm{\chi}=0$. 
Here the U(1)$_{local}^{\omega}$ part in $SU(2)_{local}^{\omega}$ is not affected by this condition. Hence, the gauge symmetry corresponding to $\bm\omega_\parallel(x)$ remains unbroken. 

Therefore, if we impose the condition (\ref{MAGcond}) to the CFN-Yang--Mills theory,
we have a gauge theory (Yang--Mills theory II) with the local gauge symmetry 
$SU(2)_{local}^{\omega=\theta}$ corresponding to the gauge transformation parameter $\bm\omega(x)=(\bm\omega_\parallel(x),\bm\omega_\perp(x)=\bm\theta_\perp(x))$, which is a diagonal SU(2) part of the original 
$SU(2)_{local}^{\omega} \times [SU(2)/U(1)]_{local}^{\theta}$.  
The local gauge symmetry $SU(2)_{local}^{\omega=\theta}$  of the Yang--Mills theory II is the same as the gauge symmetry II.


\section{BRST symmetry and Faddeev--Popov ghost term}


According to the clarification of the symmetry in the CFN-Yang--Mills theory explained above,  we can obtain the unique Faddeev-Popov ghost term associated to the gauge fixing adopted in quantization.  
This is  an advantage of our interpretation of the CFN-Yang--Mills theory. 


\subsection{Determination of BRST transformations}


We introduce two kinds of ghosts, $\mathbb C_\omega$ and $\mathbb C_\theta$ corresponding to ${\bm\omega}$ and ${\bm\theta}$, respectively. 
Then, by requiring 
\begin{align}
 {\bm n}\cdot\mathbb C_\theta=0 ,
\end{align}
the BRST transformations for 
 $\mathscr A_\mu$ and $\bm n$ are determined as follows. 
\begin{align}
 \bm\delta_{\rm B}\mathscr A_\mu
 =D_\mu[\mathscr A]\mathbb C_\omega, \quad
\bm\delta_{\rm B}\bm n
  =g\bm n\times\mathbb C_\theta.
\end{align}


\subsubsection{$\omega$ sector}


By imposing the nilpotency for $\mathscr A_\mu$, i.e., 
$\bm\delta_{\rm B}^2 \mathscr A_\mu \equiv 0$, we can determine the BRST transformation for $\mathbb C_\omega$:
\begin{equation}
\bm\delta_{\rm B}\mathbb C_\omega
  =-\frac g2\mathbb C_\omega\times\mathbb C_\omega.
\end{equation}
as in the usual case. 
The nilpotency for $\mathbb C_\omega$, i.e., $\bm\delta_{\rm B}^2 \mathbb C_\omega \equiv 0$ can be checked in the same way as in the usual  case. 

In the similar way to the ordinary case, we can introduce the antighost $\bar{\mathbb C}_\omega$ and the Nakanishi-Lautrup (NL) auxiliary field $\mathbb N_\omega$ obeying the BRST transformations:
\begin{align}
\bm\delta_{\rm B}\bar{\mathbb C}_\omega
  =i\mathbb N_\omega, \quad
\bm\delta_{\rm B}\mathbb N_\omega
  =0.
\end{align}
The nilpotency for $\bar{\mathbb C}_\omega$ and $\mathbb N_\omega$ is trivial, i.e., $\bm\delta_{\rm B}^2 \bar{\mathbb C}_\omega \equiv 0$ and $\bm\delta_{\rm B}^2 \mathbb N_\omega \equiv 0$. 


\subsubsection{$\theta$ sector}


First, imposing the nilpotency for $\bm n$ yields a relationship, 
\begin{align}
0 \equiv \bm\delta_{\rm B}^2\bm n
=g\bm n\times\bm\delta_{\rm B}\mathbb C_\theta ,
\label{eq:delta^2n}
\end{align}
where we have used $\bm n \cdot \mathbb C_\theta=0$ and $\mathbb C_\theta \cdot \mathbb C_\theta=0$.
Second, requiring $\bm\delta_{\rm B}(\bm n\cdot\mathbb C_\theta)=0$, i.e.,  
\begin{align}
0 \equiv  \bm\delta_{\rm B}(\bm n\cdot\mathbb C_\theta)
 =\bm n\cdot(g\mathbb C_\theta\times\mathbb C_\theta
              +\bm\delta_{\rm B}\mathbb C_\theta) ,
\label{eq:delta(nC)}
\end{align}
leads to another relationship,
\begin{equation}
\bm\delta_{\rm B}\mathbb C_\theta
  =-g\mathbb C_\theta\times\mathbb C_\theta+\bm f_\perp,
  \label{eq:temp}
\end{equation}
where $\bm f_\perp$ denotes an arbitrary function perpendicular to $\bm n$. 
 Substituting the relation (\ref{eq:temp}) into (\ref{eq:delta^2n}), we have
\begin{align}
\bm n\times\bm\delta_{\rm B}\mathbb C_\theta
 =\bm n\times\bm f_\perp
 =0 ,
\end{align}
 implying $\bm f_\perp=0$. 
Hence the BRST transformation for $\mathbb C_\theta$ is determined as
\begin{equation}
\bm\delta_{\rm B}\mathbb C_\theta
  =-g\mathbb C_\theta\times\mathbb C_\theta ,
  \label{eq:BCtheta}
\end{equation}
where $\mathbb C_\theta\times\mathbb C_\theta$ is parallel to $\bm{n}$ due to (\ref{eq:delta^2n}) and (\ref{eq:BCtheta}). 
\footnote{
Alternatively, (\ref{eq:BCtheta}) is shown as follows. 
The decomposition of the BRST transformation of $\mathbb C_\theta$ into the parallel and perpendicular parts to $\bm n$
yields 
\begin{align}
\bm\delta_{\rm B}\mathbb C_\theta
 =(\bm n\cdot\bm\delta_{\rm B}\mathbb C_\theta)\bm n
   +(\bm n\times\bm\delta_{\rm B}\mathbb C_\theta)\times n
 =-g\{\bm n\cdot(\mathbb C_\theta\times\mathbb C_\theta)\}\bm n
 &=-g\mathbb C_\theta\times\mathbb C_\theta,
\end{align}
where we have used (\ref{eq:delta^2n}) and (\ref{eq:delta(nC)}), i.e.,  
$
\bm n\cdot\bm\delta_{\rm B}\mathbb C_\theta
 =-g\bm n\cdot(\mathbb C_\theta\times\mathbb C_\theta),
$
in the second equality, 
and we have used the fact that $\mathbb C_\theta\times\mathbb C_\theta$ is parallel to $\bm n$ in the last step. 
}
The nilpotency $\bm\delta_{\rm B}^2 \mathbb C_\theta \equiv 0$ can be checked without difficulty.

We introduce the antighost $\bar{\mathbb C}_\theta$ and the NL field $N_\theta$ such that they have the BRST transformations:
\begin{align}
\bm\delta_{\rm B}\bar{\mathbb C}_\theta
  =i\mathbb N_\theta, \quad
\bm\delta_{\rm B}\mathbb N_\theta
  =0.
\end{align}
The nilpotency for $\bar{\mathbb C}_\theta$ and $\mathbb N_\theta$ is trivial, i.e., 
$\bm\delta_{\rm B}^2 \bar{\mathbb C}_\theta \equiv 0$ and $\bm\delta_{\rm B}^2 \mathbb N_\theta \equiv 0$.

We have not yet imposed any conditions on 
$\bar{\mathbb C}_\theta$, $\mathbb N_\theta$, although we imposed 
$\bm n\cdot\mathbb C_\theta=0$ on $\mathbb C_\theta$.
Note that $\bar{\mathbb C}_\theta$ and  $\mathbb N_\theta$ are not necessarily perpendicular to $\bm n$. 
In light of the fact that $\mathbb C_\theta$ has two degrees of freedom, 
we impose 
\begin{align}
 \bm n\cdot\bar{\mathbb C}_\theta=0
\end{align}
and 
$\bm\delta_{\rm B}(\bm n\cdot\bar{\mathbb C}_\theta) \equiv 0$.  
Then we find 
\begin{align}
0 \equiv \bm\delta_{\rm B}(\bm n\cdot\bar{\mathbb C}_\theta)
 =\bm n\cdot(g\mathbb C_\theta\times\bar{\mathbb C}_\theta
              +\bm\delta_{\rm B}\bar{\mathbb C}_\theta)
 =\bm n\cdot(g\mathbb C_\theta\times\bar{\mathbb C}_\theta
              +i\mathbb N_\theta) .
\end{align}
This condition implies that the parallel component $\mathbb N_\theta^{\parallel}$ is nonzero 
and written in terms of $\mathbb C_\theta$, $\bar{\mathbb C}_\theta$ and $\bm n$:
\begin{equation}
\mathbb N_\theta^{\parallel}
 :=\bm n(\bm n\cdot\mathbb N_\theta)
  =ig\bm n[\bm n\cdot(\mathbb C_\theta\times\bar{\mathbb C}_\theta)]  
  =ig (\mathbb C_\theta\times\bar{\mathbb C}_\theta)  ,
\end{equation}
where we have used a fact that $\mathbb C_\theta\times\bar{\mathbb C}_\theta$ is parallel to $\bm{n}$, since ${\bm n}\cdot\mathbb C_\theta=0=0=\bm n\cdot\bar{\mathbb C}_\theta$. 
Therefore, $\mathbb N_\theta^{\parallel}$ is not the independent degree of freedom.


\subsubsection{Summarizing BRST transformations}


Thus we have determined the nilpotent BRST transformations for the CFN variables based on a new interpretation \cite{KMS05} of the CFN-Yang--Mills theory. 

The CFN variables obey the BRST transformations:
\begin{subequations}
\begin{align}
\bm\delta_{\rm B}\bm n
  =& g\bm n\times\mathbb C_\theta^\perp ,\\
  \bm\delta_{\rm B} c_\mu  
  =&  g(\bm{n} \times \mathscr{A}_\mu ) \cdot (\mathbb C_\omega^\perp - \mathbb C_\theta^\perp ) + \bm{n} \cdot \partial_\mu \mathbb C_\omega ,
  \label{btc}
\\
  \bm\delta_{\rm B} \mathbb{X}_\mu 
  =& g \mathbb{X}_\mu \times  (\mathbb C_\omega^\parallel +\mathbb C_\theta^\perp ) + D_\mu[\mathbb{V}](\mathbb C_\omega^\perp -\mathbb C_\theta^\perp ) ,
  \label{btX}
\end{align}
which are supplemented by the BRST transformations in the $\omega$ sector
\begin{align}
\bm\delta_{\rm B}\mathbb C_\omega
 &=-\frac g2\mathbb C_\omega\times\mathbb C_\omega,\\
  \bm\delta_{\rm B}\bar{\mathbb C}_\omega
 &=i\mathbb N_\omega,\\
\bm\delta_{\rm B}\mathbb N_\omega
 &=0 ,
\end{align}
and the BRST transformations in the $\theta$ sector
\begin{align}
\bm\delta_{\rm B}\mathbb C_\theta^\perp
 &=-g\mathbb C_\theta^\perp \times\mathbb C_\theta^\perp ,\\
\bm\delta_{\rm B}\bar{\mathbb C}_\theta^\perp &=i\mathbb N_\theta^\perp 
- g \mathbb C_\theta^\perp \times\bar{\mathbb C}_\theta^\perp ,\\
 \bm\delta_{\rm B}\mathbb N_\theta^\perp
&=  g\mathbb N_\theta^\perp\times\mathbb C_\theta
 - g^2 i(\mathbb C_\theta\cdot\bar{\mathbb C}_\theta) \mathbb C_\theta 
\nonumber\\
&=  g(\mathbb N_\theta^\perp -ig\mathbb C_\theta\times\bar{\mathbb C}_\theta) \times \mathbb C_\theta 
= - \bm\delta_{\rm B} \mathbb N_\theta^{\parallel} , 
\label{btNt}
\end{align}
\end{subequations}
where we have explicitly written the BRST transformation of the perpendicular $\mathbb N_\theta^\perp$ component in conjunction to the  parallel $\mathbb N_\theta^{\parallel}$ component.
\footnote{The (\ref{btNt}) is obtained from the nilpotency for $\bar{\mathbb C}_\theta$:
\begin{align}
0 \equiv \bm\delta_{\rm B}^2\bar{\mathbb C}_\theta
 =i\bm\delta_{\rm B}\mathbb N_\theta
=i\bm\delta_{\rm B}(\mathbb N_\theta^\perp+\mathbb N_\theta^{\parallel})
 &=i\bm\delta_{\rm B}
    (\mathbb N_\theta^\perp+ig\mathbb C_\theta\times\bar{\mathbb C}_\theta)
   \nonumber\\
 &=i\bm\delta_{\rm B}
    \mathbb N_\theta^\perp
   +ig\mathbb C_\theta\times
    \mathbb N_\theta^\perp
    - g^2(\mathbb C_\theta\cdot\bar{\mathbb C}_\theta)
        \mathbb C_\theta .
\end{align}
}
Here $\mathbb C_\theta^\perp$, $\bar{\mathbb C}_\theta^\perp$ and $\mathbb N_\theta^\perp$ have two independent degrees perpendicular to $\bm n$.  


\subsection{Gauge-fixing and the FP ghost term}


The gauge-fixing (GF) and the associated Faddeev--Popov (FP) ghost term is written as follows. 
\begin{subequations}
\begin{align}
{\cal L}_{\rm GF+FP}
 =&{\cal L}_{\rm GF+FP}^\omega  +{\cal L}_{\rm GF+FP}^\theta,
\\
&  {\cal L}_{\rm GF+FP}^\omega
 := -i\bm\delta_{\rm B}(\bar{\mathbb C}_\omega\cdot\mathbb F_\omega)
 =-i\bm\delta_{\rm B}(\bar{\mathbb C}_\omega\cdot\partial^\mu\mathscr A_\mu),
\\
& {\cal L}_{\rm GF+FP}^\theta
 := -i\bm\delta_{\rm B}(\bar{\mathbb C}_\theta\cdot\mathbb F_\theta)
 =-i\bm\delta_{\rm B}
     (\bar{\mathbb C}_\theta\cdot D^\mu[\mathbb V]\mathbb X_\mu) .
\end{align}
\end{subequations}
The first term 
${\cal L}_{\rm GF+FP}^\omega$ is calculated in the similar way to the ordinary   Lorentz gauge as 
\begin{align}
{\cal L}_{\rm GF+FP}^\omega
=\mathbb N_\omega\cdot\partial^\mu \mathscr A_\mu
   +i\bar{\mathbb C}_\omega\cdot
    \partial^\mu D_\mu[\mathscr A]\mathbb C_\omega,
\label{eq:L_omega}
\end{align}

In calculating the second term ${\cal L}_{\rm GF+FP}^\theta$, a useful observation is that the new MAG condition $\mathbb F_\theta =D^\mu[\mathbb V]\mathbb X_\mu$ 
can be rewritten in terms of $\mathscr{A}_\mu$ and $\bm{n}$ as
\begin{align}
\mathbb F_\theta
 &=D^\mu[\mathbb V]\mathbb X_\mu
  =D^\mu[\mathscr A]\mathbb X_\mu
=g^{-1}D^\mu[\mathscr A]\mathbb(\bm n\times D_\mu[\mathscr A]\bm n)
   \nonumber\\
 &=g^{-1}\bm n\times D^\mu[\mathscr A]D_\mu[\mathscr A]\bm n.
\end{align}
Then, it is straightforward but a little bit tedious to show that 
\begin{align}
{\cal L}_{\rm GF+FP}^\theta
 =\mathbb N_\theta^\perp\cdot(D^\mu[\mathbb V]\mathbb X_\mu)
  -i\bar{\mathbb C}_\theta\cdot
    D^\mu[\mathbb V-\mathbb X]D_\mu[\mathbb V+\mathbb X]
    (\mathbb C_\theta-\mathbb C_\omega) ,
\label{eq:L_theta2}
\end{align}
where we have used that 
$\bar{\mathbb C_\theta}\cdot\bm n=0=\mathbb C_\theta\cdot\bm n$ and
$\bm n\cdot D_\mu[\mathscr A]\bm n=0$ . 
This is one of the main results of this paper. 
Note that ${\cal L}_{\rm GF+FP}^\theta$ includes the ghost field $\mathbb C_\omega$ which can not be eliminated  from (\ref{eq:L_theta2}) by shifting the variable as 
$\mathbb C_\theta\rightarrow\mathbb C_\theta-\mathbb C_\omega$.
This is because $\mathbb C_\omega$ have three degrees, while 
$\mathbb C_\theta$ have two degrees.

In the one-loop calculation, however, we can eliminate $\mathbb C_\omega$ by shifting $\mathbb C_\theta\rightarrow\mathbb C_\theta-\mathbb C_\omega^\perp$, if  we treat $\mathbb V$ as a background and $\mathbb X$, $\mathbb C$, $\bar{\mathbb C}$ as the quantum fluctuation around it:
\begin{align}
{\cal L}_{\rm GF+FP}^\theta
 &=
  \mathbb N_\theta^\perp\cdot(D^\mu[\mathbb V]\mathbb X_\mu)
  -i\bar{\mathbb C}_\theta\cdot
    D^\mu[\mathbb V]D_\mu[\mathbb V]
    (\mathbb C_\theta-\mathbb C_\omega)
   +\cdots
   \nonumber\\
 &=
  \mathbb N_\theta^\perp\cdot(D^\mu[\mathbb V]\mathbb X_\mu)
  -i\bar{\mathbb C}_\theta\cdot
    D^\mu[\mathbb V]D_\mu[\mathbb V]
    (\mathbb C_\theta-\mathbb C_\omega^\perp)
   +\cdots
   \nonumber\\
 &=
  \mathbb N_\theta^\perp\cdot(D^\mu[\mathbb V]\mathbb X_\mu)
  -i\bar{\mathbb C}_\theta\cdot
    D^\mu[\mathbb V]D_\mu[\mathbb V]
    \mathbb C_\theta
   +\cdots
\end{align}
Thus, the previous result in the one-loop level \cite{Kondo04} is not affected by the correct treatment of the FP ghost term. 

Moreover, observing 
$
\bm\delta_{\rm B}(\mathbb N_\omega\cdot\mathbb N_\omega)=0,
\quad
\bm\delta_{\rm B}(\mathbb N_\theta \cdot \mathbb N_\theta)=0,
$
and
$
\bm\delta_{\rm B}(\mathbb N_\theta +\zeta\mathbb N_\omega )^2
=0 ,
$
we are allowed to add the following term to the GF+FP term. 
\begin{equation}
{\cal L}_N
 :=\frac\alpha2
    (\mathbb N_\omega + \zeta \mathbb N_\theta)
    \cdot(\mathbb N_\omega + \zeta \mathbb N_\theta)
   +\frac\beta2\mathbb N_\omega\cdot\mathbb N_\omega .
   \label{newNN}
\end{equation}
The usefulness of this term is demonstrated in the modified new MAG in the final part of this paper.


\section{Primed gauge and BRST transformations}



\subsection{Primed gauge transformations}


Another way to describe the gauge transformation property of the CFN-Yang--Mills theory in terms of the CFN variables $(\bm n,c_\mu,\mathbb X_\mu)$ is to introduce 
\begin{equation}
\bm\omega^\prime:=\bm\omega_{\parallel}+\bm\theta_\perp,
\quad
\bm\theta^\prime:=\bm\omega_\perp-\bm\theta_\perp,
\quad
(\bm\theta^\prime\cdot\bm n=0) , 
\end{equation}
rather than $\bm\omega$ and $\bm\theta$.  
The relationship between two sets of gauge transformation parameters:
\begin{subequations}
\begin{align}
\bm\omega^\prime+\bm\theta^\prime
  =\bm\omega_{\parallel}+\bm\omega_\perp
  =\bm\omega,
 \quad 
\bm n\times(\bm\omega^\prime\times\bm n)
  =\bm\omega_\perp^\prime
  =\bm\theta_\perp
  =\bm\theta,
\end{align}
\end{subequations}
yields another view of the local gauge transformations:
\begin{subequations}
\begin{align}
\delta\mathscr A_\mu
 &=D_\mu[\mathscr A](\bm\omega^\prime+\bm\theta^\prime),
   \\
\delta\bm n
 &=g\bm n\times\bm\omega_\perp^\prime
  =g\bm n\times\bm\omega^\prime,
   \\
\delta c_\mu
 &=g(\mathscr A_\mu\times\bm n)\cdot\bm\theta^\prime
   +\bm n\cdot\partial_\mu(\bm\omega^\prime+\bm\theta^\prime),
   \\
\delta\mathbb X_\mu
 &=g\mathbb X_\mu\times\bm\omega^\prime
   +D_\mu[\mathbb V]\bm\theta^\prime .
\end{align}
\end{subequations}
In fact, $\bm\theta^\prime=0$ i.e., $\bm\omega_\perp=\bm\theta_\perp$ reproduces the local gauge transformation II.


\subsection{Primed BRST transformations}


By introducing the ghost fields 
$\mathbb C_\omega^\prime$ and 
$\mathbb C_\theta^\prime$
corresponding to 
$\bm\omega^\prime$ and 
$\bm\theta^\prime$ respectively, 
we can introduce the BRST transformations of the CFN variables $(\bm n,c_\mu,\mathbb X_\mu)$ in addition to the original Yang-Mills field $\mathscr{A}_\mu$. 
\begin{subequations}
\begin{align}
\bm\delta_{\rm B}\mathscr A_\mu
 &=D_\mu[\mathscr A](\mathbb C_\omega^\prime+\mathbb C_\theta^\prime),
   \\
\bm\delta_{\rm B}\bm n
 &=g\bm n\times\mathbb C_\omega^\prime,
   \\
\bm\delta_{\rm B}c_\mu
 &=g(\mathscr A_\mu\times\bm n)\cdot\mathbb C_\theta^\prime
   +\bm n\cdot\partial_\mu(\mathbb C_\omega^\prime+\mathbb C_\theta^\prime),
   \\
\bm\delta_{\rm B}\mathbb X_\mu
 &=g\mathbb X_\mu\times\mathbb C_\omega^\prime
   +D_\mu[\mathbb V]\mathbb C_\theta^\prime .
\end{align}
In the similar way to the above, the BRST transformations of the ghost fields,  
$\mathbb C_\omega^\prime$, $\mathbb C_\theta^\prime$,  and  antighost fields, $\bar{\mathbb C}_\omega^\prime$, $\bar{\mathbb C}_\theta^\prime$, can be determined as well as the NL fields $\mathbb N_\theta^\prime$, $\mathbb N_\omega^\prime$ as 
\begin{align}
\bm\delta_{\rm B}\mathbb C_\theta^\prime
 &=-g\mathbb C_\omega^\prime\times\mathbb C_\theta^\prime,
   \\
\bm\delta_{\rm B}\bar{\mathbb C}_\theta^\prime
 &=i\mathbb N_\theta^\prime,
   \\
\bm\delta_{\rm B}\mathbb N_\theta^\prime
 &=0,
   \\
\bm\delta_{\rm B}\mathbb C_\omega^\prime
 &=-\frac g2
    (\mathbb C_\omega^\prime\times\mathbb C_\omega^\prime
     +\mathbb C_\theta^\prime\times\mathbb C_\theta^\prime),
   \\
\bm\delta_{\rm B}\bar{\mathbb C}_\omega^\prime
 &=i\mathbb N_\omega^\prime,
   \\
\bm\delta_{\rm B}\mathbb N_\omega^\prime
 &=0 ,
\end{align}
\end{subequations}
where
$
\bm n\cdot\mathbb C_\theta^\prime
 =0
 =\bm n\cdot\bar{\mathbb C}_\theta^\prime,
\quad
\bm n\cdot\mathbb N_\theta^\prime
 =ig\bm n\cdot(\mathbb C_\theta^\prime\times\bar{\mathbb C}_\theta^\prime) .
$


\subsection{Primed gauge fixing and FP ghost term}


We impose two gauge fixing conditions: 
$
\mathbb F_\omega^\prime
 =\mathbb F_\omega
 =\partial^\mu\mathscr A_\mu 
$
for fixing the gauge degrees $\bm\omega^\prime$ 
and 
$
\mathbb F_\theta^\prime
 =\mathbb F_\theta
 =D^\mu[\mathbb V]\mathbb X_\mu  
$
for fixing gauge degrees $\bm\theta^\prime$.
Then we can obtain the primed GF+FP term in the similar way to the above,
\begin{align}
{\cal L}_{\rm GF+FP}
 =& {\cal L}_\omega^\prime
  +{\cal L}_\theta^\prime,
\nonumber\\
{\cal L}_\omega^\prime
 &:=-i\bm\delta_{\rm B}(\bar{\mathbb C}_\omega^\prime\cdot\mathbb F_\omega)
=-i\bm\delta_{\rm B}
     (\bar{\mathbb C}_\omega^\prime\cdot\partial_\mu\mathscr A_\mu)
    \nonumber\\
 &=\mathbb N_\omega^\prime\cdot\partial_\mu\mathscr A_\mu
   +i\bar{\mathbb C}_\omega^\prime
     \cdot\partial^\mu D_\mu[\mathscr A]
     (\mathbb C_\omega^\prime+\mathbb C_\theta^\prime) ,
\label{GF1}
\\
{\cal L}_\theta^\prime
 &:=-i\bm\delta_{\rm B}(\bar{\mathbb C}_\theta^\prime\cdot\mathbb F_\theta)
=-i\bm\delta_{\rm B}
     (\bar{\mathbb C}_\theta^\prime\cdot D_\mu[\mathbb V]\mathbb X_\mu)
    \nonumber\\
 &=\mathbb N_\theta^\prime\cdot D_\mu[\mathbb V]\mathbb X_\mu
   \nonumber\\
 &\quad
   +i\bar{\mathbb C}_\theta^\prime\cdot
    \{g(D_\mu[\mathbb V]\mathbb X^\mu)\times
      (\mathbb C_\omega^\prime-\mathbb C_\theta^\prime)
       +D_\mu[\mathbb V-\mathbb X]
        D_\mu[\mathbb V+\mathbb X]\mathbb C_\theta^\prime\}.
\end{align}
The second term is further rewritten into
\begin{align}
{\cal L}_\theta^\prime
 &=\mathbb N_\theta^\prime\cdot D_\mu[\mathbb V]\mathbb X_\mu
   \nonumber\\
 &\quad
   +i\bar{\mathbb C}_\theta^\prime\cdot
    \{-g(\mathbb C_\omega^\prime)_{\parallel}
       \times(D_\mu[\mathbb V]\mathbb X^\mu)
      +D_\mu[\mathbb V-\mathbb X]
        D_\mu[\mathbb V+\mathbb X]\mathbb C_\theta^\prime\}
   \nonumber\\
 &=\mathbb N_\theta^\prime\cdot D_\mu[\mathbb V]\mathbb X_\mu
   +i\bar{\mathbb C}_\theta^\prime\cdot
    D_\mu[\mathbb V-\mathbb X]D_\mu[\mathbb V+\mathbb X]
     \{\mathbb C_\theta^\prime+(\mathbb C_\omega^\prime)_{\parallel}\} .
\label{GF2}
\end{align}
Note that (\ref{GF1}) and (\ref{GF2}) agree with (\ref{eq:L_omega}) and (\ref{eq:L_theta2}) under the identification:
\begin{align}
\mathbb C_\omega
  =\mathbb C_\omega^\prime+\mathbb C_\theta^\prime, \quad
\mathbb C_\theta
  =(\mathbb C_\omega^\prime)_\perp.
\end{align}
This is expected from the observation that the correspondence between the original parameters and ghosts 
$\bm\omega\rightarrow\mathbb C_\omega$,
$\bm\theta\rightarrow\mathbb C_\theta$,
and the primed parameters and ghosts 
$\bm\omega^\prime\rightarrow\mathbb C_\omega^\prime$,
$\bm\theta^\prime\rightarrow\mathbb C_\theta^\prime$. 
This can be a cross check for the correctness of our calculations.


\section{Anti-BRST transformation and its application} %


By the formal replacement $\mathbb C\rightarrow\bar{\mathbb C}$,
$\bar{\mathbb C}\rightarrow\mathbb C$  and
$\mathbb N\rightarrow\bar{\mathbb N}$, 
we begin to determine the anti-BRST transformation. 


\subsection{Anti-BRST transformation} %


The anti-BRST transformations for $\omega$ sector are obtained as
\begin{subequations}
\begin{align}
\bar{\bm\delta}_{\rm B}\mathscr A_\mu
 &=D_\mu[\mathscr A]\bar{\mathbb C}_\omega,\\
\bar{\bm\delta}_{\rm B}\bar{\mathbb C}_\omega
 &=-\frac g2\bar{\mathbb C}_\omega\times\bar{\mathbb C}_\omega,\\
\bar{\bm\delta}_{\rm B}\mathbb C_\omega
 &=i\bar{\mathbb N}_\omega,\\
\bar{\bm\delta}_{\rm B}\bar{\mathbb N}_\omega
 &=0.
\end{align}
\end{subequations}
We require $\{\bm\delta_{\rm B},\bar{\bm\delta}_{\rm B}\}=0$ to obtain the relationship between $\mathbb{N}$ and $\bar{\mathbb{N}}$.
For $\mathscr A_\mu$, using  
$
\bm\delta_{\rm B}\bar{\bm\delta}_{\rm B}\mathscr A_\mu
=\bm\delta_{\rm B}
   (D_\mu[\mathscr A]\bar{\mathbb C}_\omega)
=iD_\mu[\mathscr A]\mathbb N_\omega
   +gD_\mu[\mathscr A]\mathbb C_\omega\times\bar{\mathbb C}_\omega,
$
we obtain 
\begin{align}
0 \equiv (\bm\delta_{\rm B}\bar{\bm\delta}_{\rm B}
 +\bar{\bm\delta}_{\rm B}\bm\delta_{\rm B})\mathscr A_\mu
 &=iD_\mu[\mathscr A]
   (\mathbb N_\omega
    +\bar{\mathbb N}_\omega
    -ig\mathbb C_\omega\times\bar{\mathbb C}_\omega),
\end{align}
yielding the relationship 
\begin{equation}
\mathbb N_\omega+\bar{\mathbb N}_\omega
 =ig\mathbb C_\omega\times\bar{\mathbb C}_\omega .
\end{equation}
On the other hand, the anti-BRST transformations for the $\theta$ sector are obtained as
\begin{subequations}
\begin{align}
\bar{\bm\delta}_{\rm B}\bm n
 &=g\bm n\times\bar{\mathbb C}_\theta^\perp,\\
\bar{\bm\delta}_{\rm B}\bar{\mathbb C}_\theta^\perp
 &=-g\bar{\mathbb C}_\theta^\perp\times\bar{\mathbb C}_\theta^\perp,\\
\bar{\bm\delta}_{\rm B}\mathbb C_\theta^\perp
 &=i\bar{\mathbb N}_\theta^\perp,\\
\bar{\bm\delta}_{\rm B}\bar{\mathbb N}_\theta^\perp 
 &= 0 ,
\end{align}
\end{subequations}
where $\bar{\mathbb N}_\theta$ have two independent degrees $\bar{\mathbb N}_\theta^\perp$, since 
\begin{equation}
\mathbb N_\theta^{\parallel}
 =ig\bm n[\bm n\cdot(\bar{\mathbb C}_\theta^\perp\times\mathbb C_\theta^\perp)]
 =ig\bar{\mathbb C}_\theta^\perp\times\mathbb C_\theta^\perp.
\end{equation}

For $\bm n$, we calculate
$
\bm\delta_{\rm B}\bar{\bm\delta}_{\rm B}\bm n
=\bm\delta_{\rm B}
   (g\bm n\times\bar{\mathbb C}_\theta^\perp)
=-g^2\bm n(\mathbb C_\theta^\perp\cdot\bar{\mathbb C}_\theta^\perp)
   +ig\bm n\times\mathbb N_\theta,
$
 and hence
\begin{align}
0 \equiv (\bm\delta_{\rm B}\bar{\bm\delta}_{\rm B}
 +\bar{\bm\delta}_{\rm B}\bm\delta_{\rm B})\bm n
 &=ig\bm n\times(\mathbb N_\theta+\bar{\mathbb N}_\theta),
\end{align}
which implies the relationship
\begin{equation}
\mathbb N_\theta^\perp+\bar{\mathbb N}_\theta^\perp
 =0 , 
\quad 
\mathbb N_\theta+\bar{\mathbb N}_\theta
 =
\mathbb N_\theta^{\parallel}+\bar{\mathbb N}_\theta^{\parallel}
 =2ig\mathbb C_\theta^\perp\times\bar{\mathbb C}_\theta^\perp .
\end{equation}
We have checked that the requirement 
$\bm\delta_{\rm B}\bar{\bm\delta}_{\rm B}+\bar{\bm\delta}_{\rm B}\bm\delta_{\rm B} \equiv 0$ 
leads to no new relationships among the fields. 


\subsection{Modified new MAG}


As an application of the anti-BRST transformation, we propose a modified version \cite{KondoII} of the new MAG as 
\begin{align}
S_{GF+FP} &= \int d^D x i\bm\delta_{\rm B}\bar{\bm\delta}_{\rm B}
\left(\textstyle\frac12\mathbb X_\mu\cdot\mathbb X^\mu\right)
   \nonumber\\
 &=\int d^D x \,
   i^{-1}\bm\delta_{\rm B}
     [(\bar{\mathbb C}_\omega-\bar{\mathbb C}_\theta^\perp)
      \cdot D^\mu[\mathbb V]\mathbb X_\mu]
   \nonumber\\
 &=\int d^D x \Big\{ 
(\mathbb N_\omega^\perp-\mathbb N_\theta^\perp)
     \cdot(D^\mu[\mathbb V]\mathbb X_\mu)
   \nonumber\\
 &\qquad\qquad
   -i(\bar{\mathbb C}_\omega^\perp-\bar{\mathbb C}_\theta^\perp)\cdot
     D^\mu[\mathbb V-\mathbb X]D_\mu[\mathbb V+\mathbb X]
     (\mathbb C_\theta^\perp-\mathbb C_\omega) \Big\} .
\end{align}
This term is exact simultaneously in the BRST and anti-BRST transformation and invariant under the FP conjugation. 
This form of the FP term can be cast into a simplified form by taking an appropriate linear combination of two NL fields, $\mathbb N_\omega^\perp$ and $\mathbb N_\theta^\perp$,  in (\ref{newNN}) and by integrating out the NL fields.

\section{Conclusion}

In this paper we have determined the BRST and anti-BRST transformations of the CFN variables and those of the associated ghost, antighost and Nakanishi-Lautrup auxiliary fields. 
The general form of the gauge-fixing term for the new MAG and the correct form of the associated FP term are obtained explicitly based on the new interpretation \cite{KMS05} of the CFN-Yang-Mills theory.  
Although the general form is different from the previous one \cite{Kondo04},   it reduces in the one-loop level  to the same form as that given in \cite{Kondo04} and does not change the main result \cite{Kondo04} obtained based on the one-loop expression. 
Finally, a modified form of the new MAG has been proposed using the BRST and anti-BRST transformations.

\section*{Acknowledgments}
This work is supported by 
Grant-in-Aid for Scientific Research (C)14540243 from Japan Society for the Promotion of Science (JSPS), 
and in part by Grant-in-Aid for Scientific Research on Priority Areas (B)13135203 from
the Ministry of Education, Culture, Sports, Science and Technology (MEXT).


\end{document}